\documentclass[doublecol,figures]{epl2}
\usepackage{amsmath}
\usepackage{amsfonts}
\usepackage{amssymb}

\title{Helical, Angular and Radial Ordering \\in Narrow Capillaries}
\shorttitle{Ordering in Capillaries}

\author{I. Erukhimovich\inst{1} \and A. Johner\inst{2}}
\shortauthor{I. Erukhimovich \etal}

\institute{
  \inst{1} Moscow State University, Moscow
119992 Russia\\
  \inst{2} Institute Charles Sadron, 6 rue Boussingault, 67083 Strasbourg
Cedex, France }
\pacs{nn.mm.xx}{First pacs description}
\pacs{nn.mm.xx}{Second pacs description} \pacs{nn.mm.xx}{Third
pacs description}

\abstract{To enlighten the nature of the order-disorder and
order-order transitions in block copolymer melts confined in
narrow capillaries we analyze peculiarities of the conventional
Landau weak crystallization theory of systems confined to
cylindrical geometry. This phenomenological approach provides a
quantitative classification of the cylindrical ordered
morphologies by expansion of the order parameter spatial
distribution into the eigenfunctions of the Laplace operator. The
symmetry of the resulting ordered morphologies is shown to
strongly depend both on the boundary conditions (wall preference)
and dimensionless parameter $q_{*}R$, where $R$ is the cylinder
radius and $q_{*}$ is the wave number of the critical order
parameter fluctuations, which determine the bulk ordering of the
system under consideration. In particular, occurrence of the
helical morphologies is a rather general consequence of the
imposed cylindrical symmetry for narrow enough capillaries. We
discuss also the ODT and OOT involving some other simplest
morphologies. The presented results are relevant also to other
ordering systems as charge-density waves appearing under addition
of an ionic solute to a solvent in its critical region, weakly
charged polyelectrolyte solutions in poor solvent, microemulsions
etc.}

\begin{document}

\maketitle

\section{Introduction}
During the last years there has been large interest in the
behavior of block copolymer (basically diblock copolymer) melts
confined in narrow capillaries, which were studied both
experimentally \cite{nmat,Science,Macro5660,Macro1055,JPSB43,MRC}
and via computer simulation \cite{JCPMC,PRL138306,Macro4899} as
well as by numerical calculations within the self-consistent field
theory \cite{nmat,Sevink2001,Macro806}. The most characteristic of
these morphologies are lamellae ordered normal to the cylinder
axis $z$ (slab morphology\cite{Sevink2001} ) and coaxial shells or
multiwall morphologies\cite{Sevink2001,JCPMC} as well as various
helices\cite{nmat,PRL138306,Macro4899}, the morphologies' symmetry
being strongly depended on the ratio of the gyration radius
$R_{G}$ of the diblock macromolecules under consideration and the
radius $R$ of the cylinder the macromolecules are confined to.
However, some important issues have not been addressed yet. In
particular, the various morphologies (up to 20 in ref
\cite{Macro806} and up to 30 in ref \cite{PRL138306}) are
distinguished only visually by their snapshots and no attempts to
relate them to a quantitative order parameter are made until now.
Meanwhile, without a quantitative definition of an order parameter
and, accordingly, finding the quantitative change of such an order
parameter at the phase transitions one cannot discuss the
transitions between the morphologies rigorously. Probably, this is
why no accurate phase diagram of the order-disorder and
order-order transitions for cylindrically confined block copolymer
melts is presented until now. In particular, the phase diagram
presented in ref \cite{Macro806} has two important shortcomings:
\emph{i}) it is based on a visual distinguishing of the
morphologies \textit{only}, and \emph{ii}) the authors assume no
modulation of the block copolymer composition along the $z$-axes,
which, obviously, excludes the slab and helical morphologies.
Besides, an alternation of some coexisting phases is to occur in
cylindrical capillaries already by virtue of the Landau
theorem\cite{Landau} as in any effective 1D system.

Another shortcoming of all the aforementioned studies
\cite{nmat,Science,Macro5660,Macro1055,JPSB43,MRC,JCPMC,PRL138306,Sevink2001,Macro806,Macro4899}
is that they are specifically block copolymer oriented whereas
ordering in cylindrically confined systems seems to be much more
general. Indeed, the very existence of the various morphologies in
block copolymer melts is due to the instability of their
homogeneous phase with respect to spatial fluctuations of the
polymer concentration with a finite spatial
period\cite{deGennes79m,Leibler,Erukh82}. Thus, the ODT in block
copolymers is only a particular case of the well known weak
crystallization instability\cite{Landau37,KLMreview}. The Landau
weak crystallization theory (WCT) provides a common physical
background for describing such seemingly different physical
phenomena as \emph{i}) forming weakly segregated morphologies in
molten block copolymers\cite{Leibler}, therewith the polymer
specific features appear at the stage of the microscopic
calculation of the Landau expansion coefficients only, \emph{ii})
appearance of the blue phases in liquid crystals \cite{BP1,BP2},
\emph{iii}) charge-density waves generation upon addition of an
ionic solute to a solvent in its critical
region\cite{Nabutov,Stell}, \emph{iv}) microphase separation in
weakly charged polyelectrolyte solutions\cite{Borue88,JL90} and
\emph{v}) microemulsions\cite{ciach}.

So, the peculiarities found for cylindrically confined block
copolymer melts are expected to exist for other nano-ordered
systems either and the aim of the present Letter is to present a
phenomenological theory of the ordering in narrow capillaries
based on the general weak crystallization paradigm.

1. We start with the standard expression for the free energy of
the weakly crystallized systems described by a non-uniform scalar
order parameter profile $\Phi(\mathbf{r}),$ which without any loss
of generality could be written as the Landau expansion in powers
of $\Phi$ up to the 4th order:
\begin{align}
\frac{vF_{\text{bulk}}}{T}  & =\int\left( \tau\,\Phi^{2}\left(
\mathbf{r}\right)  +\left(
(1+q_{\ast}^{-2}\Delta)\,\Phi(\mathbf{r})\right)
^{2}\right.  \label{free}\\
& \left.  +\alpha\Phi^{3}(\mathbf{r})+\Phi^{4}(\mathbf{r})\right)
d\mathbf{r},\nonumber
\end{align}
where $v$ is a constant having dimensionality $[l^3]$ to be
defined from microscopic considerations and $\tau$ is an effective
dimensionless temperature measured from the instability boundary
for bulk. In bulk the quadratic term in (\ref{free}) reads
\begin{equation}
F_{\text{bulk}}^{(2)}=\int\left[  \tau\,+\left(  1-\left.
q^{2}\right/  q_{\ast}^{2}\right)  ^{2}\right]  \,\left\vert \Phi_{\mathbf{q}%
}\right\vert ^{2}\frac{d\mathbf{q}}{(2\pi)^{3}} \label{free2}%
\end{equation}
where $\Phi_{\mathbf{q}}=\int\Phi\left(  \mathbf{r}\right)
\exp\left( i\mathbf{qr}\right)  \,d\,\mathbf{r}$ is the Fourier
component of the order parameter. Obviously, for $\tau>0$ a
minimum (at least, a metastable one) of the free energy
(\ref{free}) is provided for the disordered state
($\Phi(\mathbf{r})=0$), whereas for $\tau<0$ the disordered state
is absolutely unstable with respect to the growth of the Fourier
components $\Phi _{\mathbf{q}}$ with wave numbers $q=\left\vert
\mathbf{q}\right\vert $ close to the value $q=q_{\ast}$. Thus, the
parameter $q_{\ast}^{2}>0$ characterizes the wave length $L=\left.
2\pi\right/  q_{\ast}$ of the critical order parameter
fluctuations, which destroy the disordered state in bulk.

The coefficient $\alpha$ appearing in the free energy expression
(\ref{free}) depends on the physical nature of the system. In this
paper we restrict ourselves to the case $\alpha=0$, when the only
ordered phase appearing in bulk systems described by (\ref{free})
is lamellar, and show that even in this case the presence of
confinement could result in various morphologies formed by
straight and curved rod-like units.

2. In the presence of a confinement the order parameter
$\Phi(\mathbf{r})$ is to be expanded not in the Fourier harmonics,
which are the eigenfunctions of the Laplace operator in the
infinite volume, but in the eigenfunctions of the Laplace operator
\begin{equation}
\Delta\Psi\left(  \mathbf{r}\right)  =-E\Psi\left(
\mathbf{r}\right)
\label{eigen1}%
\end{equation}
for the specified confinement geometry. In the cylindrical
co-ordinates, which are the natural choice for ordering in
capillaries, the eigenvalues and the corresponding eigenfunctions
read
\begin{align}
\label{eigen} E &  =\alpha_{m,n}^{2}+p^{2},\qquad\Psi\left(
\mathbf{r}\right)
=\sum\limits_{m=-\infty}^{\infty}\sum_{n=1}^{\infty}\nonumber\\
&  \int_{-\infty}^{\infty}A_{p,m,n}\,\exp\left(  ipz\right)
\,\exp\left( im\varphi\right)  \,\phi_{m,n}\left(  r\right)
\frac{dp}{2\pi}\;%
\end{align}
Here the order parameter modulations along the cylinder axis and
in polar angle are characterized by the wave number $p$ and the
integer number $m$, respectively, whereas $\phi_{m,n}(r)$ are the
eigenfunctions of the $2$-dimensional radial Laplace operator:
\begin{equation}
\frac{d}{r\,dr}\left(  r\frac{d\phi}{dr}\right)  -\frac{m^{2}}{r^{2}}%
\phi=-\alpha_{m,n}^{2}\phi\,.\label{Bessel}%
\end{equation}
$\phi_{m,n}(r)$ are the Bessel functions\cite{Bessel} $J_{m}\left(
\alpha_{m,n}\,r\right)  $, the eigenvalues $\alpha_{m,n}^{2}$
being determined by the corresponding boundary condition. Choosing
the reflecting boundary condition
\begin{equation}
\left.  \left.  \partial\Phi\right/  \partial r\right\vert _{r=R}%
=0\,,\label{a}%
\end{equation}
which is often used to describe the behavior of confined
incompressible block copolymer melts \cite{AJS,Stepanow}, we get
$\alpha_{m,n}^{2}=s_{m,n}^{2}/R^{2}$, where $s_{m,n}$ are the
locations of extrema of the Bessel function $J_{m}\left( x\right)$
numerated in ascending order.

3. Substituting eqs (\ref{eigen}) into (\ref{free2}) we get
\begin{equation*}
F_{2}= R^{2}\int_{-\infty}^{\infty}\sum\limits_{m=-\infty
}^{\infty}\sum_{n=1}^{\infty}\Lambda_{m,n}(\tau,p,\rho)\,\mathcal{N}%
_{m,n}\,\left\vert A_{p,m,n}\right\vert ^{2}dp,
\end{equation*}
\begin{equation}
\Lambda_{m,n}(\tau,p,\rho)=\tau+(p^{2}/q_{\ast}^{2}-\kappa_{m,n}^{2}
)^{2},\quad\label{lambda}\\
\end{equation}
where $\rho=q_{\ast}R$ is the reduced capillary radius,
$\mathcal{N}_{m,n}=\int_{0}^{1}J_{m}^{2}\left( s_{m,n}\,x\right)
\,x\,dx$ and
\begin{equation}
\kappa_{m,n}^{2}=1-s_{m,n}^{2}/\rho^{2}\label{kappa}.
\end{equation}

The condition providing instability of the disordered state with
respect to infinitesimal fluctuations of the $(m,n)$-mode is
\begin{equation}
\Lambda_{m,n}(\tau,p^{\ast}_{m,n},\rho)<0, \label{spin}
\end{equation}
where $p^{\ast}_{m,n}$ is location of the absolute minimum of the
function (\ref{lambda}).

It follows from eqs (\ref{lambda}), (\ref{kappa}) that the
condition (\ref{spin}) takes two rather different forms:
\begin{equation}
p^{\ast}_{m,n}=q_{\ast}\kappa _{m,n},\quad \tau<0, \qquad
\text{for} \quad s_{m,n}<\rho \label{A}
\end{equation}
\begin{equation}
p^{\ast}_{m,n}=0, \;\quad \tau<-\kappa _{m,n}^{4} \,, \qquad
\text{for} \quad s_{m,n}>\rho. \label{B}
\end{equation}

In other words, given a finite reduced capillary radius $\rho$ all
unstable modes with $s_{m,n}>\rho$ result in an angular and radial
modulation only and their contribution to the total order
parameter profile is proportional to
$J_{m}(s_{m,n}\,r/R)\,\cos(m\varphi)$. On the contrary, the
unstable modes with $s_{m,n}<\rho$ possess both \emph{an axial}
and, generally, angular as well as radial modulation.

In the narrow capillaries obeying the inequality
\begin{equation}
\rho<s_{1,1}=1.841  \label{ineq1}
\end{equation}
the only mode $(0,1)$, which corresponds to the eigenfunction
\begin{equation}
\Phi_{0,1}\left(\mathbf{r}\right)=A_{0}\cos(q_{\ast}z)\,J_{0}\left(
s_{01}r/R\right)  =A_{0}\cos(q_{\ast}z), \label{fi0}
\end{equation}
is axially modulated ($s_{0,1}=0$). Within the temperature
interval $(-\kappa_{1,1}<\tau<0)$ minimizing the free energy
(\ref{free}) we get
\begin{equation}
A_0^2=-2\tau/3, \qquad F_{\text{bulk}}=-T (L/R)\,\tau^2/
\texttt{Gi}, \label{F0}
\end{equation}
$L$ and $\texttt{Gi}=6v/(\pi R^3)$ being the length of the
capillary and Ginzburg parameter, respectively.

\begin{figure}
\onefigure{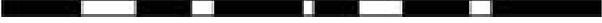} \caption{A typical succession of the ordered
(black) and disordered (white) capillary segments for a finite
value of the Ginzburg parameter.} \label{fig.1}
\end{figure}

4. Indeed, by virtue of the Landau theorem \cite{Landau}, the
ordered phase appears as an alternating succession of the ordered
and disordered segments of various, generally, lengths (see
fig.~\ref{fig.1}) rather than ordering in the whole capillary.
Accordingly, the capillary partition function could be written
similarly to that of 1d Ising (helix-coil transition) model
\cite{Huang,GK}:
\begin{equation}
Z\left( L\right) =\frac{1}{2\pi i}\int \overline{Z}(p)\,\exp
\left( pL\right) \,dp, \label{Ising1}
\end{equation}
where the Laplace transform $\overline{Z}(p)$ reads
\begin{equation*}
\overline{Z}(p)=\int_{0}^{\infty}\exp\left(  -pL\right)
\,Z(L)\,dL=\sigma _{i}\sigma_{j}Z_{ij}(p),
\end{equation*}
\begin{equation}\label{Ising2}
\Vert Z_{ij}(p)\Vert =\left(
\begin{array}
[c]{cc}
p & -\sigma\\
-\sigma & p+f
\end{array}
\right)  ^{-1},
\end{equation}
where $Tf$ is the specific (per unit length) free energy of the
ordered phase, $i,j=1,2$ and $\sigma$, $\sigma_1$, $\sigma_2$ are,
respectively, the weights to be assigned to transient zones
between the ordered and disordered segments as well as those
between the end-walls of the capillary and the disordered and
ordered segments. \footnote{\, Strictly speaking, the free energy
$F(l)$ of a capillary segment of the length $l$ is slightly
different from $fl$ since for finite segments one should use
Fourier series and $l$-dependent boundary conditions on the
end-walls rather than Fourier integrals (5b), which difference is
neglected in this paper.} It follows from eqs (\ref{Ising1}),
(\ref{Ising2}) that the length fraction of the ordered segments is
a continuous function of the parameter $u=f/2\sigma$ ($u<0$ for
$\tau<0$)
\begin{equation}
\phi_{\texttt{ord}}=-L^{-1}\partial\ln Z(L)/\partial f
=(\sqrt{u^2+1}-u)/(\,2\sqrt{u^2+1}\,) \label{fiord}
\end{equation}
Taking into account that $\sigma\sim R^{-1}\exp(fR)$ and,
therefore, $|u|\sim|fR|\exp|fR|$, we conclude that capillary
ordering is not a genuine phase transition but a co-operative one,
the location and width of the transition region being $\tau=0$ and
$|\tau|\sim(\texttt{Gi})^{1/2}$. Thus, if the inequality
(\ref{ineq1}) holds and $\texttt{Gi}\ll1$ then, as is seen from eq
(\ref{F0}), for practical purposes the line $\tau=0$ could be
considered as the 2nd order ODT line.

The presented consideration of capillary ordering is extended
straightforwardly to the case when the free energy (\ref{free})
has $N>1$ minima corresponding to different morphologies, which
provides the set $f_1,..f_N$ of the specific (per unit length)
free energies. Accordingly, the matrix $Z_{ij}$ and vector
$\sigma_i$ appearing in eq (\ref{Ising2}) become $N+1$-dimensional
ones. Then an equilibrium between the segments with different
ordered morphologies and the disordered segments would occur,
character of which would change on crossover lines. In the limit
$\texttt{Gi}\rightarrow0$ these crossover lines transform into the
1st order transition lines
\begin{equation}\label{trans}
\min(f_1,..f_N)=f_A=f_B
\end{equation}
between the ordered morphologies A and B.

5. One more mode, (1,1), is axially modulated for
\begin{equation}
s_{1,1}<\rho<s_{2,1}=3.054, \label{ineq2}
\end{equation}
two types of the axially modulated profiles being possible:
\begin{subequations}
\label{J1}%
\begin{align}
\Phi_{1,1}^{(h)}\left(  \mathbf{r}\right)   &  =A_{h}\cos\left(
\varphi\pm
p_{1,1}z\right)  \,J_{1}\left(  s_{1,1}\,r/R\right)  ,\label{h}\\
\Phi_{1,1}^{(s)}\left(  \mathbf{r}\right)   &  =A_{s}\cos\left(
\varphi\right)  \,\cos\left(  p_{1,1}z\right)  \,J_{1}\left(
s_{1,1}
\,r/R\right)  . \label{s}%
\end{align}
\end{subequations}
Further we refer to the profiles (\ref{h}) and (\ref{s}) as the
running along the axes $z$ (helical) and standing waves,
respectively. The specific free energies obtained by substituting
the order parameter profiles (\ref{J1}) into the free energy
(\ref{free}) read:
\begin{subequations}
\label{FJ1}
\begin{align}
\widetilde{f}_{h} &  =\tau \mathcal{N}_{m,n} A_{h}^{2}+
(3/4) \,\mathcal{B}_{m,n}\,A_{h}^{4}  ,\label{Fh}\\
\widetilde{f}_{s} &  =(\tau/2) \mathcal{N}_{m,n} A_{s}^{2}+
(9/32) \,\mathcal{B}_{m,n}\,A_{s}^{4}, \label{Fs}%
\end{align}
\end{subequations}
where $\widetilde{f}=fR\,\texttt{Gi}$ and
$\mathcal{B}_{m,n}=\int_{0}^{1} J_{1}^{4}(s_{m,n}\,x)\,x\,dx$,
(the difference in the corresponding numerical coefficients is due
to difference in averaging the powers of cosines (for $F_{h}$) and
the products of such powers (for $F_{s}$)). Minimizing the free
energies (\ref{FJ1}) with
respect to the corresponding amplitudes results finally in the expressions%
\begin{equation}
\widetilde{f}_{h}=-2\,\tau^{2}\mathcal{N}_{m,n}^{2}\,
\mathcal{B}_{m,n}^{-1},\quad
\widetilde{f}_{s}=-(4/3)\,\tau^{2}\mathcal{N}_{m,n}^{2}\,
\mathcal{B}_{m,n}^{-1}. \label{Fhs}%
\end{equation}
As is seen from (\ref{Fhs}), the single helical wave is always
more thermodynamically advantageous than the standing one just due
to the symmetry properties. The free energies of the left- and
right-hand helical waves are identical and the capillary would
split into the left- and right-hand helical segments with the
average segment length $\sim \sigma^{-2}$ if the helical state is
dominant.

For $\rho>s_{2,1}$ at least one more mode (2,1) becomes axially
modulated, the contribution of the mode into the total order
parameter profile resembles the double (DNA-like) helix:
\begin{equation}
\Phi_{2,1}^{(h)}\left(\mathbf{r}\right)=A_{2h}\cos\left(
2\varphi\pm p_{2,1}z\right)  \,J_{2}\left(  s_{2,1}\,r/R\right)
\label{2h}
\end{equation}

To check whether and when one non-uniform mode gives way to
another more thermodynamically advantageous one or a mixture of
two modes with further change of the temperature and capillary
radius, we substitute into the total free energy (\ref{free}) a
2-mode trial order parameter profile
\begin{eqnarray}\label{2mode}
\Phi(r) =A_1\mathcal{N}_1\cos(q_{\ast}\kappa_{m_1,n_1}z-m_1
\phi)\,J_{m_1}(s_{m_1,n_1}\,r/R)\\ \nonumber
+A_2\mathcal{N}_2\cos(q_{\ast}\kappa_{m_2,n_2}z-m_2
\phi)\,J_{m_2}(s_{m_2,n_2}\,r/R)
\end{eqnarray}

The proper choice of the normalizations $\mathcal{N}_i$ leads to
the reduced specific free energy
\begin{align}\label{2-amp}
\widetilde{f} & = 3\min \textrm{F}(A_1,A_2), \quad
\textrm{F}(A_1,A_2)=\tau_1 A_{1}^{2}+\tau_2 A_{2}^{2} \nonumber \\
& +
(3/8)(C_1c_{\,11}A_{1}^{4}+2c_{\,12}A_{1}^{2}A_{2}^{2}+C_2c_{\,22}A_{2}^{4}),
\end{align}
where $c_{ij}=2\left. \int_{0}^{1}J_{m_{i}}^{2}\left(
s_{m_{i},n_{i}}x\right) J_{m_{j}}^{2}\left(
s_{m_{j},n_{j}}x\right)xdx\right/ \mathcal{N}_{i}\mathcal{N}_{j}$,
$c_{ii}=\left. \int_{0}^{1}J_{m_{i}}^{4}\left(
s_{m_{i},n_{i}}x\right) xdx\right/\mathcal{N}_{i}^{2}$,
$\tau_{i}=\tau+\kappa_{m_i,n_i}^4$ for the modes (\ref{A}) and
$\tau_{i}=\tau$ for the modes (\ref{B}), $C_i=2/3$ if the $i$-th
mode is a mode $(0,n)$ belonging to the type~(\ref{A}) and $C_i=1$
otherwise.

The straightforward analysis shows that the minimum of the free
energy (\ref{2-amp}) is achieved in a 1-mode state only ($A_{1}=0$
or $A_{2}=0$) if the determinant $|c_{ij}|$ is negative, the
transition between the modes being the discrete 1st order phase
transition in the limit $\texttt{Gi}\rightarrow0$. If $|c_{ij}|>0$
then the minimum of the free energy (\ref{2-amp}) could correspond
to a 2-mode state ($A_{1}\neq0$ and $A_{2}\neq0$), the second mode
appearing (in this limit) via the continuous 2nd order phase
transition.

The corresponding calculations are straightforward for the
reflecting boundary condition (\ref{a}). Somewhat unexpectedly
from the mathematical standpoint there is only one dominant mode
$(0,1)$, which corresponds to the order parameter profile
$\Phi(r)=A \exp(iq_{\ast}z)$. All other modes, even those unstable
being single, are suppressed in the presence of this dominant
mode.  But from the physical viewpoint it is quite natural.
Indeed, the boundary condition (\ref{a}) does not change the
ordering conditions as compared to that in bulk, where the
lamellar morphology \emph{is} the most stable one. Thus, any other
ordered morphology would cause only some additional frustration
without any compensation.

6. The situation changes drastically if there is a preferential
adsorption of the system particles to the boundary, which is
naturally described by an additional linear surface
term\cite{ghf87}:
\begin{equation}
F= F_{\text{bulk}}+ F_{\text{surf}},\quad
F_{\text{surf}}=(hT/v)\int \Phi\left( \mathbf{r}\right)
\,dS,\label{surf}
\end{equation}
$h$ being the strength of the preferential adsorption. Obviously,
the term (\ref{surf}) affects the modes $(0,n)$ with $n$ high
enough to be of the type (\ref{B}) and thus favors the coaxial
morphologies. Say, the mode (0,2) is dominant in the narrow
capillaries satisfying the inequality
\begin{equation}
\rho<s_{0,2}=3.832  \label{ineq3}
\end{equation}
if $|\tau|$ is not too big. If $\tau\rightarrow-\infty$ the mode
$(0,1)$ becomes dominant since it has the smallest quantitative
value of the reduced forth vertex. To find the 1st order
transition line (\ref{trans}) between the modes $(0,1)$ and
$(0,2)$, it is convenient to rewrite eq (\ref{2-amp}) in the
reduced variables $A=H^{\,1/3}B$, $\tau=H^{\,2/3}\,t$,
$H^{\,4/3}\widetilde{f} = 3\min \textrm{F}(B_1,B_2)$:
\begin{align}\label{surf1}
\textrm{F}(B_1,B_2) & = b\,B+t_1B_{1}^{\;2}+t_2B_{2}^{\;2}\\
&+(3/4)B_{1}^{4}+3B_{1}^{2}B_{2}^{2}+(c_{22}/4)B_{2}^{4}.\nonumber
\end{align}
The fields $B_1,B_2$ correspond to the modes $(0,1),(0,2)$,
$b=6J_0(s_{0,2})/(\mathcal{N}_{0,2})^{1/2}$,
$\widetilde{\tau}_2=\widetilde{\tau}+\kappa_{0,2}/h^{2/3}$ and
$H=h/R$.

\begin{figure*}
\begin{center}
\includegraphics{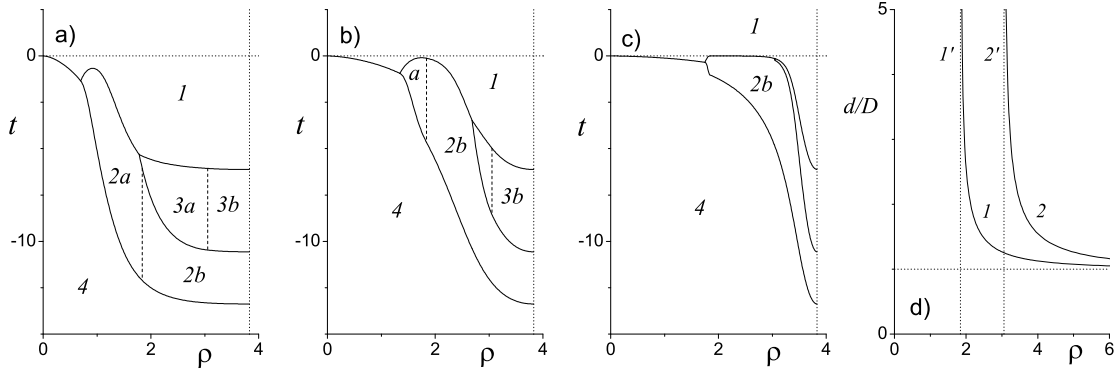}
\caption{The reduced phase diagrams of ordering in the most narrow
capillaries satisfying condition~(\ref{ineq3}) for $h=100$
(\emph{a}), $h=1$ (\emph{b}) and $h=0.01$ (\emph{c}) as well as
(\emph{d}) the dependences of the pitches $d$ (normalized on the
lamellar period in bulk $D$) of the ordinary ($1$) and double
($2$) helices on the reduced capillary radius $\rho$. The regions
$1$ and $4$ correspond to the coaxial $(0,2)$ and slab $(0,1)$
modes, the regions $2$ and $3$ do to the mixed $(0,2)+(1,1)$ and
$(0,2)+(2,1)$ states, respectively. The dotted horizontal and
vertical lines on figures (\emph{a})-(\emph{c}) correspond to the
bulk ODT temperature $\tau=0$ and the boundary~(\ref{ineq3}),
respectively. The dotted vertical lines $1'$ and $2'$ on figure
(\emph{d}) are the asymptotes $\rho=s_{1,1}$ and $\rho=s_{2,1}$ of
the curves $1$ and $2$, they correspond to the dashed vertical
lines on figures (\emph{a})-(\emph{c}) separating the regions
\emph{a} and \emph{b} of the axially non-modulated and modulated
states of the same modes. The dotted horizontal line $d=D$ on
figure (\emph{d}) is the common asymptote of the curves $1$ and
$2$ at $\rho\rightarrow\infty$.} \label{fig.2}
\end{center}
\end{figure*}

In fact, however, the transition found by minimization of the
function~(\ref{surf1}) turns our to be only metastable. The case
is that, contrary to the slab mode $(0,1)$, the coaxial mode
$(0,2)$ could be compatible (form a mixed state) with some helical
or angular modes, in particular, simple helix $(1,1)$, double
helix $(2,1)$, $(3,1),(3,2),(4,1),(4,2)$ and so on. Thus, to build
the phase diagram we are to analyze, which of the pure or mixed
modes is the most thermodynamically favorable and in which
temperature interval. In fact, it is sufficient to take into
account only the modes $\{(1,1),(2,1)\}$, which would start to
grow early than the coaxial mode $(0,2)$ in the absence of any
preferential adsorption. The resulting phase diagrams within the
interval~(\ref{ineq3}) are shown in fig.~\ref{fig.2}. Remember
that in the regions 2b, 3b the long segments with the left-hand
helical waves alternate with the right-hand ones since the free
energies of both are identical.

\begin{figure}
\onefigure{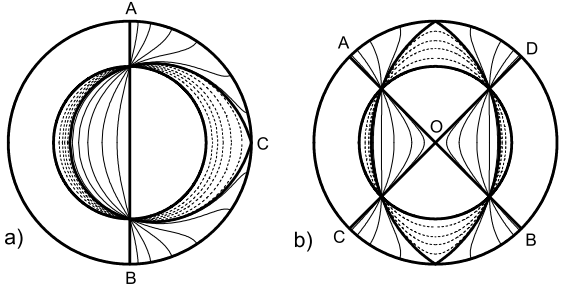} \caption{The curves (\ref{cross}) (domain
interfaces) in the 1-thread helix $(0,2)+(1,1)$  and double
DNA-like helix $(0,2)+(2,1)$. The outer bold circle is the
capillary cross-section, the inner one is the interface for the
pure $(0,2)$ mode ($x=\infty$). With decrease of $x$ the 1-thread
helix interface (\emph{a}) becomes more and more eccentric (the
plotted dashed curves correspond consecutively to
$x=5,3,2.25,1.75,1.5$) until the interface (bold solid curve)
touches the capillary wall at $x=1.445$. Next the contact area
extends (see the thin solid curves corresponding to
$x=1.4,1.2,0.8,0.5,0.25,0.1$) and at $x=0$ the interface is just
the diameter AB. The double helix interface (\emph{b}) undergoes a
similar evolution but the fact that it is centrosymmetrical.}
\label{fig.3}
\end{figure}

It is easy to see that increase of the surface field $h$ results
in increase of the temperature width $\Delta \tau$ ($\Delta
\tau\sim 10\,h^{2/3}$) of the region, where the helical ordering
can be observed. Besides, the regions of purely angular ordering
$2a$ and $3a$, which are just negligible for small $h$, grow
considerably with increase of $h$. It is convenient to visualize
the character of the appearing angular and/or helical ordering by
the curves $r(\phi)$ satisfying equation
\begin{equation}\label{cross}
\Phi(r,\phi)=0,
\end{equation}
where the order parameter $\Phi(\vect{r})$ is given by expression
~(\ref{2mode}) with $m_1=0$,\,$n_1=2$ and $m_2=1(2)$,\,$n_2=1$.
The curves~(\ref{cross}), which could be considered as
intersections of the conditional interfaces between the domains
with the cross-sections normal to the capillary axes, are plotted
in fig.~\ref{fig.3} for various values of the ratio $x=A_1/A_2$,
where $A_i$ are the coefficients appearing in the
expression~(\ref{2mode}).

Beyond the interval~(\ref{ineq3}) the mode $(0,2)$ is axially
modulated and does not contribute into the surface
term~(\ref{surf}) anymore. If $|\tau|$ is not too big, then the
modes $(0,n)$ are dominant within the intervals
$s_{0,n-1}<\rho<s_{0,n}$). With the temperature decrease the pure
coaxial modes are subsequently replaced by their various
combination with the compatible helical or angular modes until the
slab mode $(0,1)$ wins finally for $\tau\rightarrow-\infty$.
Thereby, the sets of relevant compatible modes are fast growing
with increase of the reduced capillary radius $\rho$ and, thus,
$n$, which makes calculation of the dominance maps rather
cumbersome. E.g., the relevant sets are
$\{(1,1),(2,1),(3,1),(4,1)\}$ for $n=3$ and
$\{(1,1),(2,1),(3,1),(4,1),(3,2),(4,2)\}$ for $n=4$ (other modes
with $s_{m,n}<s_{0,3(4)}$ are incompatible with the corresponding
coaxial modes). The occurring morphologies appear as ordinary,
double (DNA-like) or multi-helix structures (sometimes coaxial
ones with the same pitch) entwining central straight rod. These
morphologies resemble those seen under simulation of the
asymmetric (with non-zero cubic term) block copolymers in refs
\cite{nmat,PRL138306} and are rather different from the lamellar
morphology to be observed in the bulk for the case considered here
($\alpha=0$).

Summarizing, we have shown that emergence of various helical and
other complex morphologies in systems undergoing ordering when
they are confined to cylindrical geometry is a rather general
result of interplay between the geometrical and energetic effects
of confinement even in the seemingly trivial case of the
reflecting boundary condition~(\ref{a}) and zero cubic term. For
this case a detailed "dominance map" (in the limit
$\texttt{Gi}\rightarrow0$ it is just the phase diagram) including
various complex structures, is presented.

The actual phase diagram of particular real systems could differ
from that build in fig.~\ref{fig.2}. First, the cubic term, if
present, would contribute to stability of the found morphologies
formed by straight and curved rod-like units. Next, the numerical
values of the effective forth vertices $c_{ij}$, which appear in
the 2-amplitude Landau expansion (\ref{2-amp}) and determine
eventually the shape of the phase diagrams shown in
fig.~\ref{fig.2}, depend on the assumption of locality of the
forth vertex in the Landau Hamiltonian (\ref{free}) used in this
paper. If the vertex is non-local and reads $\int\Gamma\left(
\mathbf{r}_{1},\mathbf{r}_{2},\mathbf{r}_{3},\mathbf{r}_{4}\right)
\,\prod\limits_{i=1}^{4}\Phi\left(  \mathbf{r}_{i}\right)
\,d\mathbf{r}_{i}$, then an "angle
dependence"~\cite{Leibler,Erukh96,Erukh05} of the forth vertex
appears. When strong enough, the dependence is known to change the
sequence of the stable morphologies even in
bulk~\cite{Erukh96,Erukh05,Smirnova06}. Thus, it could change the
phase diagram shown in fig.~\ref{fig.2} either.

Our results suggest also that it could be fruitful to analyze
experimental or numerical data in terms of Bessel functions or of
their underlying symmetry. More detailed presentation of the
Landau theory including discussion of other possible boundary
conditions and microscopic theory of block copolymer will be given
in an extended version of this paper.

\acknowledgments The authors thank the program ENS-Landau for
financial support of this work.

\end{document}